\documentclass[final,5p,times,twocolumn]{elsarticle}

\journal{Optik}
\usepackage{xcolor}  
\usepackage{dashrule} 
\definecolor{verdes}{RGB}{0, 128, 0}
\definecolor{celeste}{RGB}{0,191,191}
\definecolor{rosa}{RGB}{191, 0, 191}
\usepackage{float}
\usepackage{amsmath}
\begin{document}

\title{Analytical study of optical dielectric interfaces for sensing applications}

\author[1,2,*]{German E. Caro}
\author[1]{Eduardo O. Acosta}
\author[1]{Francisco E. Veiras}
\author[1]{Liliana I. Perez}



\begin{abstract}
Dynamic optical sensors, as those based on the reflectivity of isotropic dielectrical interfaces, can be useful for measuring small changes in the refractive index of fluid media. Exact and approximated explicit formulas for their sensitivity are developed. These expressions clarify the influence of each of the constructive parameters of the device, both in partial and in total reflection. It is analytically found that sensitivities are always greater when the angle of incidence approaches the angle of total reflection. It is also demonstrated that close to the critical angle for both types of reflection, the sensitivity of the p mode is greater than that of the s mode. 

\end{abstract}


\maketitle

\section{Introduction}

Refractometry is a valuable method used in optics to determine the index of refraction of materials, which is important for assessing their composition \cite{patnaik2004dean}\cite{wilms2004determination} and purity\cite{compton2016laser}. Various techniques exist for sensing small variations in the index of refraction of a medium, including surface plasmon resonance  \cite{liang2010surface}, dielectric optical interfaces \cite{vallejo2021dielectric}, and interferometric methods \cite{rosenthal2014sensitive}. In particular \cite{zhu2017ultrasonic} presented an experimental paper based on the reflectivity of optical interfaces for ultrasound detection without providing a theoretical background to allow further optimizations. Consequently, it's possible to detect ultrasound waves by measuring small variations in refractive index. While transducers are often used for this purpose, they have some limitations in sensitivity, bandwidth, and signal-to-noise ratio \cite{wissmeyer2018looking}. On the other hand, optical methods based on refractometry offer a promising solution to these issues, and current  developments aim to improve their performance in all areas \cite{zhu2017ultrasonic} \cite{song2023toward}. These sensors can be optimized to sense small changes in the refractive index of fluids.

One of the simplest and most effective sensor schemes to sense changes in the refractive index is based on reflectivity between two linear dielectric media. Previous studies have explored the sensitivities for intensity and phase numerically \cite{vallejo2021dielectric} and experimentally 
 \cite{zhu2017ultrasonic}\cite{lin2016solution}. Numerical results indicate that the greatest sensitivities are achieved at angles close to the critical angle \cite{vallejo2021dielectric}. As far as we know, this manuscript presents the first analytical approach to this problem. Our work enables to accurately identify the conditions that result in maximum sensitivity, as detailed in the following sections.

In Section 2, we do a brief review to define the notation used in the following sections. In Section 3 we present the theoretical framework and obtain the exact analytical expressions for the sensitivity of a dielectric interface in both partial and total reflection. Moreover, we obtain approximate expressions for the sensitivity that allow us to perform a deeper analysis. In order to illustrate the importance of these expressions, in Section 4 we perform numerical analysis based on them. Finally, we present the conclusions.

\section{Reflection and refraction}

\begin{figure}
\centering
\includegraphics[width=1\linewidth]{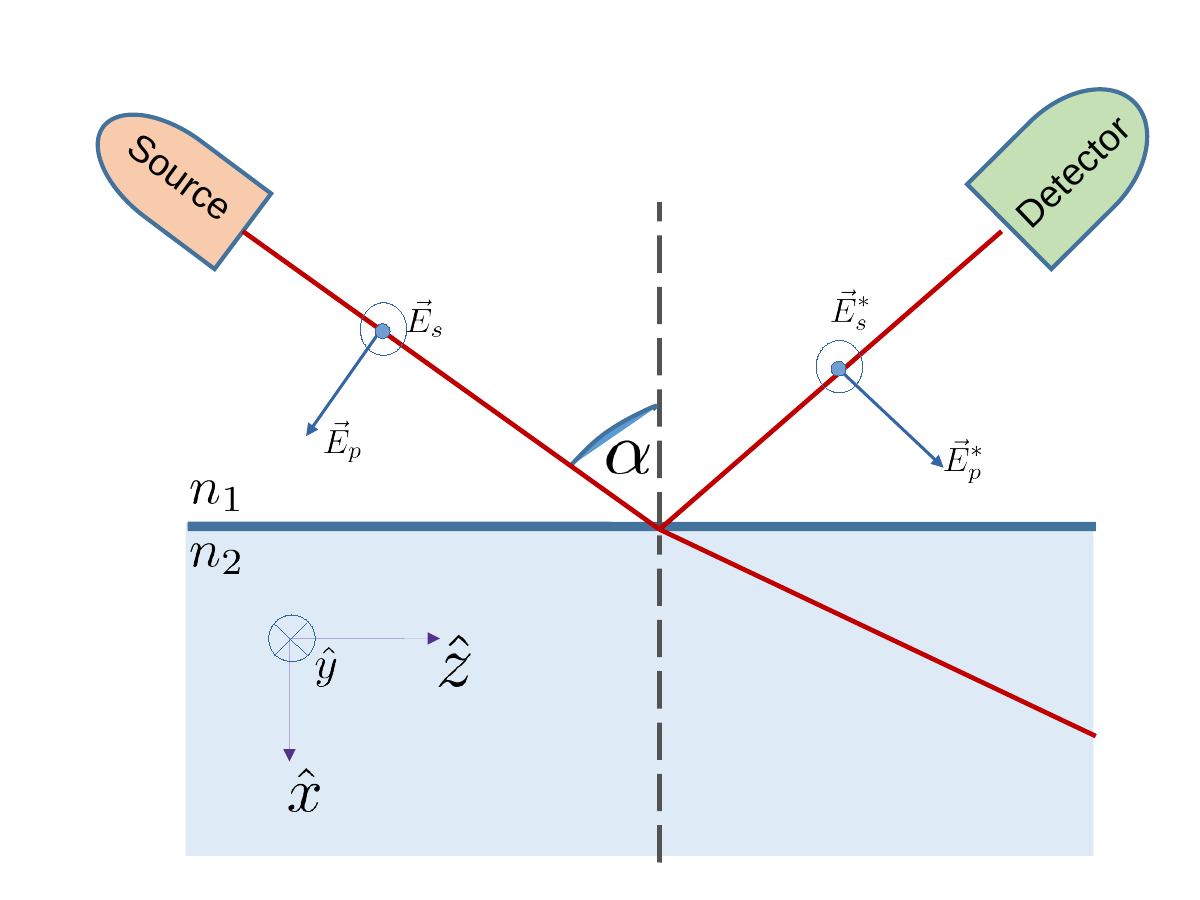}
\caption{Sensor scheme: $n_1$ and $n_2$ are the refraction indexes of the media. The light impinges on the interface with an angle of incidence $\alpha$. $\Vec{E}_s$ and $\Vec{E}_p$ are the electric fields perpendicular and parallel to the incidence plance.}
\label{fig:experimental}
\end{figure}

In this work, we consider that a monochromatic beam of known amplitude and polarization impinges with an angle $\alpha$ on the interface between two dielectric media with indexes of refraction $n_1$ and $n_2$ (Fig. \ref{fig:experimental}).

For these kinds of sensors, the relevant magnitude is the reflection coefficient. For an interface between two linear isotropic media,  the reflection coefficients are functions of $n_1$, $n_2$ and $\alpha$ and depend on the polarization modes TE ($s$) and TM ($p$) \cite{born2013principles}\cite{lekner2016theory}
\begin{equation}
\label{eq:Rs}
R_s = \frac{n_1\cos\alpha-\sqrt{n_2^2-n_1^2\sin^2\alpha}}{n_1\cos\alpha+\sqrt{n_2^2-n_1^2\sin^2\alpha}}
\end{equation}

\begin{equation}
\label{eq:Rp}
R_p = \frac{\frac{n_1}{n_2} \sqrt{n_2^2 - n_1^2\ sin^2\alpha }-n_2 \cos\alpha}{\frac{n_1}{n_2} \sqrt{n_2^2 - n_1^2 \sin^2\alpha }+n_2 \cos\alpha}
\end{equation}

From Eqs. \ref{eq:Rs} and \ref{eq:Rp} there is an angle $\alpha_c$ if $n_1 > n_2$ that makes the value inside the square root negative. 
\begin{equation}\label{eq:angulocritico}
    \sin\alpha_c = n_2/n_1 
\end{equation}

\noindent For angles of incidence smaller than the critical angle, Eqs. \ref{eq:Rs} and \ref{eq:Rp} are real, with a value between $-1$ and $1$. Nevertheless, for angles greater than the critical angle, the coefficients are complex numbers with moduli $1$ and the phases are functions of $\alpha$, $n_1$, $n_2$ and the polarization mode. In this last condition, there is no transmitted wave into the second medium.
\begin{figure}[H]
\centering
\includegraphics[width=1\linewidth]{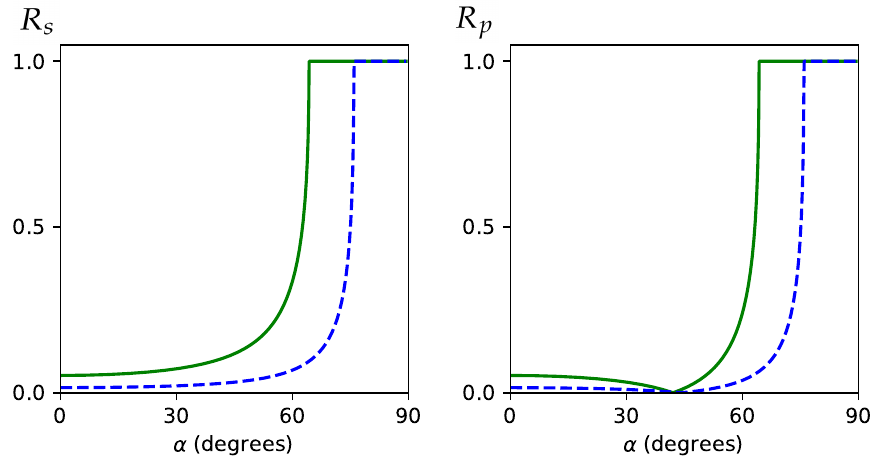}
\caption{Moduli of Fresnel coefficients for $s$ and $p$ modes as a function of $\alpha$ for $n_1=1.4865$ and two different values of $n_2$: water ({\color{verdes} \hdashrule[0.5ex]{0.3cm}{0.3mm}{}} 1.3330) and water-glicerine 75\% ({\color{blue} \hdashrule[0.5ex]{0.4cm}{0.3mm}{0.1cm 0.05cm}}1.4353).}
\label{fig:reflectividad_modulo}
\end{figure}

Fig. \ref{fig:reflectividad_modulo} shows the moduli of the reflection coefficients for $s$ and $p$ modes as a function of $\alpha$ with $\lambda=589.3$nm (D line). FK51A glass was chosen as the incident medium ($n_1=1.4865$), and two values for the refractive index of the second medium: water $n_2=1.3330$ (useful when dealing with medical applications, since it's a value close to the one from biological tissue \cite{jacques2013optical}) and a mix of water-glicerine ($25\%-75\%$) with $n_2=1.4353$ \cite{rumble2017crc}.

Total internal reflection occurs when the angle of incidence is greater than the critical angle and there is no transmitted field. Then, the phase of the reflection coefficients $\phi_s$ and $\phi_p$ are given by \cite{peatross2010physics}

\begin{equation}\label{eq:phis}
\tan\phi_s = \frac{-2n_1\cos\alpha\sqrt{n_1^2\sin^2\alpha-n_2^2}}{n_1^2-n_2^2-2(n_1^2\sin^2\alpha-n_2^2)}
\end{equation}

\noindent and 

\begin{equation}\label{eq:phip}
    \tan\phi_p=\frac{2n_1n_2^2\cos\alpha\sqrt{n_1^2\sin^2\alpha-n_2^2}}{n_2^2(n_2^2+n_1^2)-\sin^2\alpha(n_2^4+n_1^2)}
\end{equation}
\begin{figure}[H]
	\centering
	\includegraphics[width=1\linewidth]{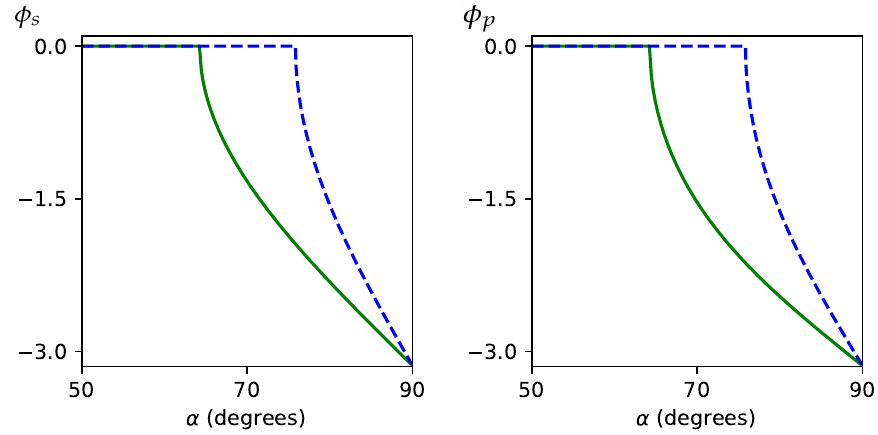}
	\caption{Phase of the Fresnel for $s$ and $p$ modes coefficients for the indexes considered in Fig. \ref{fig:reflectividad_modulo}, as a function of $\alpha$  for $n_1=1.4865$ and $n_2:$=1.3330 ({\color{verdes} \hdashrule[0.5ex]{0.3cm}{0.3mm}{}}) and $n_2$=1.4353 ({\color{blue} \hdashrule[0.5ex]{0.4cm}{0.3mm}{0.1cm 0.05cm}}). }
	\label{fig:reflectividad_fase}
\end{figure}

The phases of the Fresnel coefficients for the $s$ and $p$ modes are drawn in Fig. \ref{fig:reflectividad_fase}. In both Figs. \ref{fig:reflectividad_modulo} and \ref{fig:reflectividad_fase}, the slope presents rapid changes for angles of incidence close to the critical angle ($\alpha_c= 63.63^\circ$ for water and $\alpha_c=74.92^\circ$ for water-glicerine mixture). 

\section{Sensitivity}
The proposed sensors aim to dynamically detect small changes in the refractive index of the second medium. The changes in $n_2$ can be studied by analyzing the variations in the amplitude or phase of the reflected beam. When developing this kind of sensor, it's necessary to understand under which experimental conditions the detector can measure variations of $n_2$ with the greatest sensitivity.
From Section 2, it can be observed that for incidence angles close to the critical angle, both the reflected field in amplitude and phase exhibit a strong dependence on the angle of incidence and the refractive index of the second medium, while keeping the refractive index of the first medium constant.

When designing this type of sensors, it is necessary to study how sensitive the geometry is to changes in $\Delta n_2$. It is important to determine the optimal angle of incidence and polarization to detect small variations in $n_2$ ($\Delta n_2$).

For a fixed angle of incidence and $n_1$, the relative change in the detected magnitude is related to the sensitivity $S$ and the change in refractive index $\Delta n_2$ according to the sensitivity definition proposed by \cite{zhu2017ultrasonic} for this type of sensors:

\begin{equation}\label{eq:definicionsensibilidad}
\frac{\Delta I}{I_0} = S \Delta n_2
\end{equation}

Here, $I_0$ represents the intensity of the incident beam, and $\Delta I$ is the change in the detected intensity. The sensitivity $S$ is evaluated at a constant $n_1$ and angle of incidence ($\alpha$), as well as the mean value of $n_2$.

In the case of small variations ($\Delta n_2$), a good estimation of the sensitivity can be obtained from:

\begin{equation}\label{eq:definicionsensibilidad}
S = \frac{1}{I_0} \frac{\partial I}{\partial n_2}
\end{equation}

Here, $\frac{\partial I}{\partial n_2}$ is evaluated with the parameters $n_1$ and $\alpha$ at the mean value of $n_2$ ($\bar{n}_2$).

The sensitivity $S$ depends on the type of reflection (partial or total), polarization mode ($s$ or $p$), and the specific experimental setup employed. In the following subsections, explicit expressions are presented for each of the possible conditions, and the dependence on the parameters $n_1$, $n_2$, and $\alpha$ is analyzed.

\subsection{Partial reflection}

In the case of partial reflection, the scheme is based on sensing the intensity variations $\Delta I$ of the reflected beam for each mode of polarization. For angles of incidence smaller than the critical angle, the intensity of the reflected beam is

\begin{equation}\label{eq:ii0}
    I = I_0 \, r_{s,p}
\end{equation}

\noindent where $r_{s,p}$ are the reflectivities of $s$ and $p$ modes, respectively. Namely
\begin{equation}\label{eq:rr}
    r_{s,p} = |R_{s,p}|^2
\end{equation}


\noindent Since $\Delta I$ can be expressed in terms of the variation of the index $n_2$

\begin{equation}\label{eq:deltai}
    \Delta I = I_0 \frac{d r_{s,p}}{d n_2} \Delta n_2 
\end{equation}
\noindent  the sensitivity for each mode in partial reflection is given by  
\begin{equation}\label{eq:Samp} 
    S_{s,p}^r = \frac{dr_{s,p}}{dn_2}
\end{equation}
By means of Eqs. \ref{eq:Rs},\ref{eq:Rp}, \ref{eq:rr} and \ref{eq:Samp} the sensitivity for both modes in partial reflection can be written explicitly as

\begin{align}
    S_{s}^r =& - \frac{4 n_1 \bar{n}_2 \cos\alpha}{\sqrt{\bar{n}_2^2-n_1^2\sin^2\alpha} } 
    \frac{n_1 \cos\alpha - \sqrt{\bar{n}_2^2-n_1^2\sin^2\alpha}}{ \left(n_1 \cos\alpha + \sqrt{\bar{n}_2^2-n_1^2\sin^2\alpha}\right)^{3} } \label{eq:ss} \\
    S_{p}^r=&-\frac{4n_1 \bar{n}_2 \cos\alpha \left(2n_1^{2}\sin^{2}\alpha-\bar{n}_2^{2}\right)}{\sqrt{\bar{n}_2^2-n_1^2\sin^2\alpha}} \cdot \nonumber \\ 
    & \hspace{6em}\cdot \frac{\bar{n}_2^{2}\cos\alpha-n_1\sqrt{\bar{n}_2^2-n_1^2\sin^2\alpha}}{\left(\bar{n}_2^{2}\cos\alpha + n_1\sqrt{\bar{n}_2^2-n_1^2\sin^2\alpha}\right)^{3}} \label{eq:sp}
\end{align}

Analysing Eqs. \ref{eq:ss} and \ref{eq:sp}, it's seen that both expressions for the sensitivity have a pole in $\bar{\alpha}_C$., calculated for the indexes $n_1$ and $\bar{n}_2$. While the sensitivity for the $s$ mode is always negative and never zero, the sensitivity for the $p$ mode has two zeroes far from the critical angle.

\begin{figure}[H]
	\centering
	\includegraphics[width=1\linewidth]{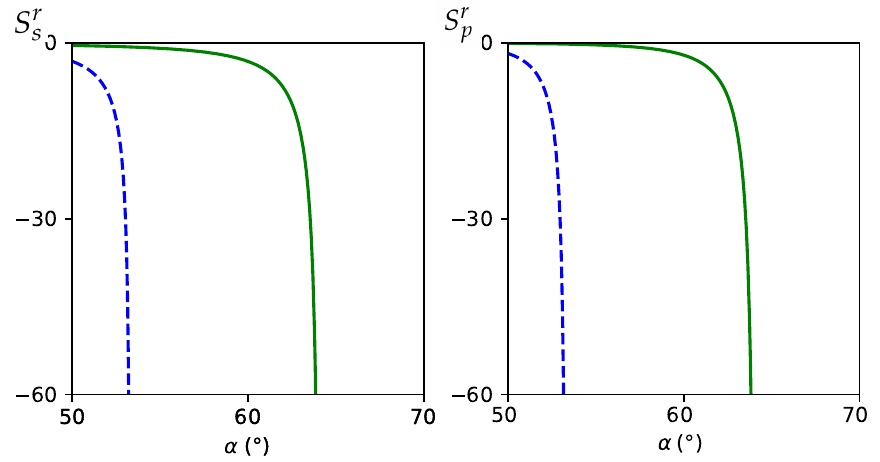}
	\caption{$S_s^r$ and $S_p^r$ as a function of $\alpha$. $\bar{n}_2=1.3330$ and two values corresponding to different commercial glasses $n_1$=1.4865 ({\color{verdes} \hdashrule[0.5ex]{0.3cm}{0.3mm}{}}) with $\bar{\alpha}_C=63.73^\circ$ and $n_1$=1.6699 ({\color{blue} \hdashrule[0.5ex]{0.5cm}{0.3mm}{0.1cm 0.05cm}}) with $\bar{\alpha}_C=52.96^\circ$.}
	\label{fig:sensibilidad_alpha}
\end{figure}

Fig. \ref{fig:sensibilidad_alpha} shows the behaviour of $S_s^r$ and $S_p^r$ for two sensors in partial reflection. Two types of glasses were considered: FK51A ($n_1=1.4835$) and BAF10 ($n_1=1.6699$). In both cases, the second medium is water. The figure was made in a range around $\bar{\alpha}_C$ for both glasses to highlight the divergence of $S_s^r$ and $S_p^r$. 


Although Eqs. \ref{eq:ss} and \ref{eq:sp} are simple enough to be used in daily calculations, it's not straightforward to extract from them which are the conditions that lead to greater sensitivities. 
Since Fig. \ref{fig:sensibilidad_alpha} shows $|S^r|$ is greater near the critical angle, it was decided to make a Taylor expansion around $\bar{\alpha}_C$ in Eqs. \ref{eq:ss} and \ref{eq:sp}.
This way, the sensitivities expressions around the critical angle can be written in a very simple and clear way (see Appendix \ref{ap:analitico})

\begin{equation}\label{eq:ssapprox}
 S_{s}^r\approx-\frac{2\sqrt{2}\sqrt{\bar{n}_2}}{(n_{1}^{2}-\bar{n}_2^{2})^{3/4}\sqrt{(\alpha_{c}-\alpha)}}+\frac{16\bar{n}_2}{(n_{1}^{2}-\bar{n}_2^{2})}
\end{equation}

\begin{equation}\label{eq:spapprox}
  S_{p}^r\approx-\frac{2\sqrt{2}n_{1}^{2}}{\bar{n}_2^{3/2}(n_{1}^{2}-\bar{n}_2^{2})^{3/4}\sqrt{(\bar{\alpha}_C-\alpha)}}+\frac{16n_{1}^{4}}{\bar{n}_2^{3}(n_{1}^{2}-\bar{n}_2^{2})}
\end{equation}

Near the critical angle, when the independent term of Eqs. \ref{eq:ssapprox} and \ref{eq:spapprox} can be ignored, the quotient $S_p^r/S_s^r=n_1^2/\bar{n}_2^2$ is greater than one. Thus, around the critical angle, the $p$ mode has greater sensitivity than the $s$ mode. 
It's clearly seen that the sensitivity has a vertical asymptote for both modes at $\alpha=\bar{\alpha}_C$ and is higher when $n_1$ approaches to $\bar{n}_2$. 


The approximations obtained are valid for angles of incidence close to $\bar{\alpha}_C$. For example, in Table 1 the distances to the critical angles are shown if  differences of $10\%$ between approximated and exact sensitivities are desired.

\begin{table}[H]\label{tab:table1}\centering
\begin{tabular}{|c|c|c|}
\hline
$n_1$  & s mode  &  p mode\\ \hline
1.4865 & 0.131                   & 0.086                  \\ \hline
1.6699 & 0.200                   & 0.083                  \\ \hline
\end{tabular}
\caption{$|\alpha-\bar{\alpha}_C|$ (degrees) for $\bar{n}_2=1.3330$ and the values for $n_1$ used in Fig. 4 if a difference of $10\%$ between approximated and exact sensitivities is desired.}
\end{table}
\subsection{Total internal reflection} 

When the incidence angle is greater than $\bar{\alpha}_C$, the reflected beam has a phase difference with regard to the incident beam. Information about this phase can be obtained by interferometric methods. 

In conditions of total reflection, the incident beam is split into two beams with the same intensity $I_0/2$ and phase difference $\phi_0$. One of them is used as a reference beam and the other impinges on the interface. The interference pattern between the reference and reflected beam is,
\begin{equation}
	I = I_0 \cos^2\left(\frac{\phi_{s,p}+\phi_0}{2}\right)
\end{equation}
where $\phi_{s,p}$ is the phase difference between the incident and the reflected beams. As in partial reflection, the dependence of $\Delta I$ on $\Delta n_2$ can be obtained
\begin{equation}
\Delta I = -I_0 \sin\left(\phi_{s,p}+\phi_0\right)\frac{d\phi_{s,p}}{d\bar{n}_2}\Delta n_2
\end{equation}


Interferometers can be adjusted to their maximum response when $|\sin\left(\phi_{s,p}+\phi_0\right)|=1$. In this condition
\begin{equation}
    \Delta I/I_0 = -\frac{d\phi_{s,p}}{d \bar{n}_2} \Delta n_2
\end{equation}
Therefore, from Eq. \ref{eq:definicionsensibilidad}, the sensitivity for total internal reflection sensors is
\begin{equation}\label{eq:sensibilidadfase}
    S_{s,p}^{\phi} =  -\frac{d\phi_{s,p}}{d n_2}
\end{equation}
Applying Eq. \ref{eq:sensibilidadfase} to Eqs. \ref{eq:phis}, \ref{eq:phip}, explicit expressions for the sensitivi\-ties for the $s$ and $p$ modes are obtained.
\begin{equation}\label{eq:ssfase}
    S_s^{\phi} = -\frac{2 n_1 \bar{n}_2 \cos\alpha}{\left(n_1^{2} - \bar{n}_2^{2}\right) \sqrt{n_1^{2} \sin^{2}\alpha - \bar{n}_2^{2}}}
\end{equation}
\begin{equation}\label{eq:spfase}
S_{p}^{\phi}=-\frac{2 n_1 \bar{n}_2\left(2 n_1^{2} \sin^{2}\alpha - \bar{n}_2^{2}\right) \cos\alpha}{\sqrt{n_1^{2} \sin^{2}\alpha - \bar{n}_2^{2}} \left(n_1^{4} \sin^{2}\alpha - n_1^{2} \bar{n}_2^{2} + \bar{n}_2^{4} \cos^{2}\alpha\right)}
\end{equation}

\begin{figure}[H]
	\centering
        	\includegraphics[width=1\linewidth]{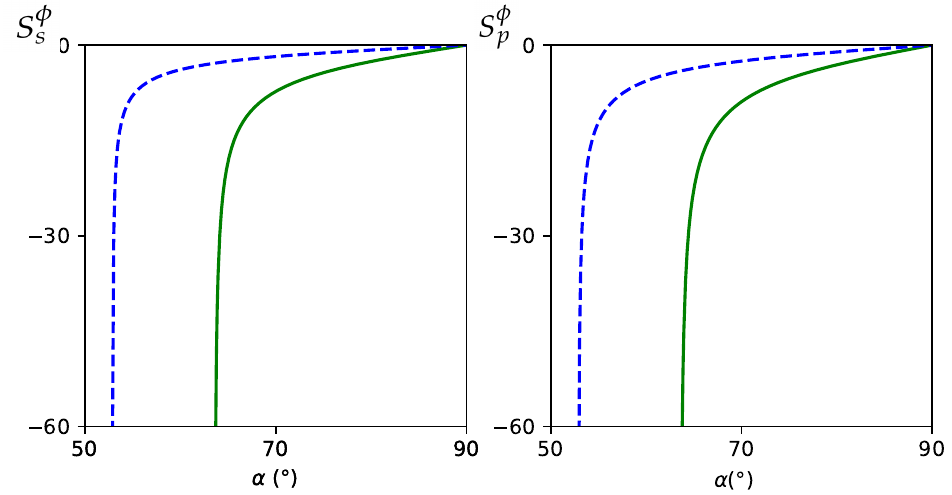}
	\caption{$S_s^\phi$ and $S_p^\phi$ as a function of $\alpha$. $\bar{n}_2=1.3330$ and two values corresponding to different commercial glasses $n_1$=1.4865 ({\color{verdes} \hdashrule[0.5ex]{0.3cm}{0.3mm}{}}) with $\bar{\alpha}_C=63.73^\circ$ and $n_1$=1.6699 ({\color{blue} \hdashrule[0.5ex]{0.5cm}{0.3mm}{0.1cm 0.05cm}}) with $\bar{\alpha}_C=52.96^\circ$.}
	\label{fig:sensibilidad_alphafase}
\end{figure}

\noindent Eqs. \ref{eq:ssfase} and \ref{eq:spfase} show there is no value of $\alpha$ such that $S^{\phi}=0$. It should be noted that Eq.\ref{eq:spfase} can be nullified for an angle outside the range of total reflection. Fig. \ref{fig:sensibilidad_alphafase} shows that $|S^\phi|$ for the $s$ and $p$ modes grow for angles of incidence close to the critical angle corresponding to each medium. Both modes have an unique vertical asymptote at $\alpha=\bar{\alpha}_C$. This behaviour is similar to that for partial reflection. 

The behaviour of the sensitivity close to the critical angle can be analysed, taking into account the Taylor series around $\alpha-\bar{\alpha}_C$ at first order (see Appendix \ref{ap:analiticofase})

\begin{equation}\label{eq:sspapprox}
    S_{s}^{\phi}\approx-\frac{\sqrt{2}\sqrt{\bar{n}_2}}{(n_{1}^{2}-\bar{n}_2^{2})^{3/4}}\frac{1}{\sqrt{\alpha-\bar{\alpha}_C}} 
\end{equation}
\begin{equation}\label{eq:sppapprox}
S_{p}^{\phi}\approx-\frac{\sqrt{2}n_{1}^{2}}{\bar{n}_2^{3/2}(n_{1}^{2}-\bar{n}_2^{2})^{3/4}}\frac{1}{\sqrt{\alpha-\bar{\alpha}_C}}
\end{equation}
As in partial reflection $S_p^\phi/S_s^\phi=n_1^2/\bar{n}_2^2$, greater than one. The sensitivity has a vertical asymptote for both modes at $\alpha=\bar{\alpha}_C$ and is greater when $n_1$ approaches $\bar{n}_2$. 

\begin{table}[H]\centering
\begin{tabular}{|c|c|c|}
\hline
$n_1$  &  s mode &  p mode \\ \hline
1.4865 & 3.093                    & 2.108 \\ \hline
1.6699 & 4.311                    & 2.108 \\ \hline
\end{tabular}
\caption{$|\alpha-\bar{\alpha}_C|$ (degrees) for $\bar{n}_2=1.3330$ and the values for $n_1$ used in Fig.  5 if a difference of $10\%$ between approximated and exact sensitivities is desired.} \label{tab:tabla2B}
\end{table} 

As in partial reflection, the approximations are good for angles of incidence close to $\bar{\alpha}_C$. For example, in Table \ref{tab:tabla2B} the distances to the critical angles are shown if differences of $10\%$ between Eqs.\ref{eq:ssfase} and \ref{eq:sspapprox}  and between Eqs. \ref{eq:spfase} and \ref{eq:sppapprox}.

\section{Analysis}
Expressions \ref{eq:ssapprox} and \ref{eq:spapprox} allow analysing the conditions to be fulfilled at the moment of the design of the experiment to sense changes in $\bar{n}_2$ in partial reflection. Analogously, Eqs. \ref{eq:sspapprox} and \ref{eq:sppapprox} are the equivalent equations for total internal reflection.

Since the $s$ and $p$ modes are qualitatively similar, as shown in Figs. \ref{fig:sensibilidad_alpha} and \ref{fig:sensibilidad_alphafase}. Moreover, near the critical angle both sensitivities are proportional, and the $p$ mode is slightly superior. Consequently, only the $p$ mode will be analysed.

\begin{figure}[H]
\centering
\includegraphics[width=1\linewidth]{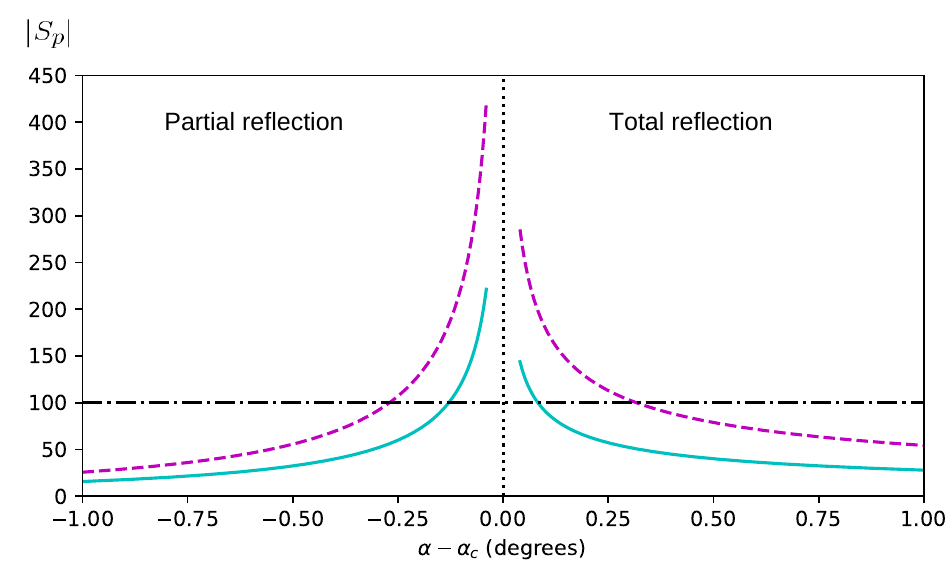}
\caption{Sensitivity for $p$ mode as a function of $\alpha-\bar{\alpha}_C$ for water $\bar{n}_2=1.3330$ ({\color{celeste} \hdashrule[0.5ex]{0.3cm}{0.3mm}{}}) and water-glicerine 75\% $\bar{n}_2=1.4353$ ({\color{rosa} \hdashrule[0.5ex]{0.5cm}{0.3mm}{0.075cm 0.05cm}} with $n_1=1.4865$).}\label{fig:SSS}
\end{figure}

A good way to compare different designs (geometry and materials) is to graph $|S_p|$ as a function of $\alpha-\bar{\alpha}_C$ instead of $\alpha$. This allows to compare the sensitivity in partial reflection and total reflection, plotting both cases in the same figure. Fig. \ref{fig:SSS} shows the sensitivity for this type of sensor for the $p$ mode, using fixed $n_1=1.4865$ and different values for $\bar{n}_2$ ($\bar{n}_2=1.3330$ and $1.4353$). Since it diverges at $\bar{\alpha}_C$, the graphs are made up to $0.04^o$ from $\bar{\alpha}_C$.
\noindent In partial reflection, the sensitivities are higher. Also, the closer $\bar{n}_2$ is to $n_1$, the greater the sensitivity, both in partial reflection and in total reflection. 
In an experimental setup, the intensity is the value to be measured in partial reflection. On the other hand, the phase is the measured magnitude in total reflection by interferometric methods. If these magnitudes are detected with a relative uncertainty of 1/1000 and the variations of $\bar{n}_2$ are of the order $1/100000$, then the setup requires a minimum sensitivity of $100$.
Table \ref{tab:tabla3} shows the results for two values of $\bar{n}_2$ to get a sensitivity of 100. For example, if $n_1=1.4865$ and $\bar{n}_2=1.4353$, the angle of incidence has to be $0.270^\circ$ for partial reflection. On the other hand, for total reflection, the angle of incidence has to be $0.136^\circ$ for $p$ mode. This again shows that the sensitivity is greater for partial reflection.
\begin{table}[H]\centering
\begin{tabular}{|c|c|c|}
\hline
$\bar{n}_2$  & Partial reflection & Total reflection \\ \hline
1.3330  & 0.129  & 0.083  \\ \hline
1.4353  & 0.270  & 0.136  \\ \hline
\end{tabular}
\caption{$|\alpha-\bar{\alpha}_C|$ (degrees) for $n_1=1.4865$ and the values of Fig. \ref{fig:SSS} when $|S_p^\phi|=|S_p^r|=100$.} \label{tab:tabla3}
\end{table}
\section{Conclusions}
This work studies the sensitivity of a sensor based on an interface between two isotropic dielectric media to measure small variations in the refractive index of the refracting medium. An analytical expression for the sensitivity is derived, providing a valuable tool for predicting its performance under different design parameters.

A first analysis reveals that the sensitivity is maximized for angles of incidence near the critical angle of total internal reflection, regardless of whether the reflection is partial or total. It is also found that the p-polarized mode is more sensitive than the s-polarized mode in all cases, which is consistent with our prior numerical research in this area. In addition, it is observed that the sensor's sensitivity is higher in partial reflection than in total reflection. This conclusion is supported by both graphical and analytical evidence.

To clarify the functional dependence of the sensor's sensitivity near the critical angle, an approximate expression is developed. These expressions provide a clear and intuitive visualization. The approximation turns out to be more accurate in total reflection than in partial reflection.

Moreover, we also found that the refractive indices of the mediums provide higher sensitivities when $\bar{n}_2$ y closer to $n_1$. This is also enhanced when both indices are large.  However, as the indices of refraction become closer, the critical angle also becomes larger. 

By means of the expressions derived, this study provides valuable insights into the design and performance of this type of sensor. 

\appendix
\section{Approximation in partial reflection}\label{ap:analitico}

Both expressions can be written as a product of two factors: only one of them diverges for $\alpha=\bar{\alpha}_C$. The other factor that goes to zero is approximated at first order. The sensitivity for $s$ and $p$ modes (Eqs. \ref{eq:ss} and \ref{eq:sp}) can be rewritten
\begin{equation}\label{eq:spu}
S_{p}=\frac{1}{u} g(u)
\end{equation}
\begin{equation}\label{eq:ssu}
    S_{s}=\frac{1}{u} h(u)
\end{equation}
where
\begin{equation}\label{eq:uu}
u=\sqrt{\bar{n}_2^{2}-n_{1}^{2}\sin^{2}\alpha}
\end{equation}
is a parameter that approaches zero when the incidence angle gets closer to the critical angle and where 
\begin{align}
g(u)=(2u^{2}-\bar{n}_2^{2})\sqrt{u^{2}+n_{1}^{2}-\bar{n}_2^{2}} \cdot \hspace{6em} \\
\hspace{3em}\cdot \frac{4n_{1}^{2}\bar{n}_2\left(n_{1}^{2}u-\bar{n}_2^{2}\sqrt{u^{2}+n_{1}^{2}-\bar{n}_2^{2}}\right)}{\left(n_{1}^{2}u+\bar{n}_2^{2}\sqrt{u^{2}+n_{1}^{2}-\bar{n}_2^{2}}\right)^{3}}  
\end{align}
and
\begin{equation}
h(u)=4\bar{n}_2\frac{\sqrt{u^2+n_{1}^{2}-\bar{n}_2^{2}}\left(u-\sqrt{u^2+n_{1}^{2}-\bar{n}_2^{2}}\right)}{\left(\sqrt{u^2+n_{1}^{2}-\bar{n}_2^{2}}+u\right)^{3}}
\end{equation}

Since $g(u)$ y $h(u)$ are finite when $u$ approaches zero, it's possible to do a first order Taylor expansion around $u=0$ such that
\begin{equation}\label{eq:guap}
    g(u)\approx \left(g(0)+\frac{\partial g}{\partial u}(0)u\right)
\end{equation}
\noindent and analogously for $h$. Where

\begin{equation}\label{eq:gcero}
    g(0)=-\frac{4n_{1}^{2}}{\bar{n}_2\sqrt{n_{1}^{2}-\bar{n}_2^{2}}}
\end{equation}
\begin{equation}\label{eq:hcero}
    h(0) = -\frac{4\bar{n}_2}{\sqrt{n_{1}^{2}-\bar{n}_2^{2}}}
\end{equation}
and
\begin{equation}\label{eq:gpcero}
    \frac{\partial g}{\partial u}(0)=-\frac{16n_{1}^{4}}{\bar{n}_2^{3}(n_{1}^{2}-\bar{n}_2^{2})}
\end{equation}
\begin{equation}\label{eq:hpcero}
    \frac{\partial h}{\partial u}(0)=-\frac{16\bar{n}_2}{(n_{1}^{2}-\bar{n}_2^{2})}
\end{equation}

Since $\sin^2\alpha$ can be approximated as 
\begin{equation}
    \sin^{2}\alpha\approx\frac{\bar{n}_2^{2}}{n_{1}^{2}}+2\frac{\bar{n}_2}{n_{1}}\sqrt{1-\frac{\bar{n}_2^{2}}{n_{1}^{2}}}(\alpha-\bar{\alpha}_C)
\end{equation}

replacing into Eq. \ref{eq:uu}
\begin{equation}\label{eq:uap}
    u\approx \sqrt{2\bar{n}_2\sqrt{n_{1}^{2}-\bar{n}_2^{2}}(\bar{\alpha}_C-\alpha)}
\end{equation}

In consequence of Eqs. \ref{eq:guap}, \ref{eq:gcero}, \ref{eq:hcero}, \ref{eq:gpcero}, \ref{eq:hpcero}, \ref{eq:uap} an approximated expression for the sensitivity as a function of the closeness to the critical angle was obtained.
\begin{equation}
 S_{s}^r\approx-\frac{2\sqrt{2}\sqrt{\bar{n}_2}}{(n_{1}^{2}-\bar{n}_2^{2})^{3/4}\sqrt{(\bar{\alpha}_C-\alpha)}}+\frac{16\bar{n}_2}{(n_{1}^{2}-\bar{n}_2^{2})}
\end{equation}

\begin{equation}
  S_{p}^r\approx-\frac{2\sqrt{2}n_{1}^{2}}{\bar{n}_2^{3/2}(n_{1}^{2}-\bar{n}_2^{2})^{3/4}\sqrt{(\bar{\alpha}_C-\alpha)}}+\frac{16n_{1}^{4}}{\bar{n}_2^{3}(n_{1}^{2}-\bar{n}_2^{2})}
\end{equation}
\section{Approximation in total reflection}\label{ap:analiticofase}
As was the case with the exact expression for the sensitivity in reflectivity, in the case of the sensitivity in phase it's not straightforward to visualize which parameters or combination allows obtaining the maximum sensitivity. Using the same techniques that in Appendix \ref{ap:analitico} to approximate these expressions close to $\bar{\alpha}_C$,
the sensitivity in phase for the $s$ and $p$ modes (Eqs. \ref{eq:ssfase} and \ref{eq:spfase}) can be rewritten as

\begin{equation}
S_p^{\phi}=\frac{1}{v}g^{\phi}(u)
\end{equation}

\begin{equation}
S_s^{\phi}=\frac{1}{v}h^{\phi}(u)
\end{equation}
where the auxiliary functions $g^{\phi}$ y $h^{\phi}$ are
\begin{equation}
g^{\phi}=-\frac{2\bar{n}_2n_{1}^{2}\left(2v{^{2}}+\bar{n}_2^{2}\right)\sqrt{n_{1}^{2}-v{^{2}}-\bar{n}_2^{2}}}{n_{1}^{4}v{^{2}}+\bar{n}_2^{4}\left(n_{1}^{2}-v{^{2}}-\bar{n}_2^{2}\right)}
\end{equation}
\begin{equation}
h^{\phi}=-\frac{2\bar{n}_2\sqrt{n_{1}^{2}-v{{}^2}+\bar{n}_2^{2}}}{\left(n_{1}^{2}-\bar{n}_2^{2}\right)}
\end{equation}
calling $v$
\begin{equation}
    v=\sqrt{n_{1}^{2}\sin^{2}\alpha-\bar{n}_2^{2}}
\end{equation}
Since $g^{\phi}$ and $h^{\phi}$ are finite when $v$ approach zero, it's possible to take a Taylor expansion of first order around $v=0$ such that
\begin{equation}
   g^{\phi}(v) \approx \left(g^{\phi}(0)+\frac{\partial g^{\phi}}{\partial u} (0) u\right)
\end{equation}
and analogously for $h$. where
\begin{equation}
g^{\phi}(0)=-\frac{2n_{1}^{2}}{\bar{n}_2\sqrt{n_{1}^{2}-\bar{n}_2^{2}}}
\end{equation}
\begin{equation}
h^{\phi}(0) = -\frac{2\bar{n}_2}{\sqrt{n_{1}^{2}-\bar{n}_2^{2}}}
\end{equation}
Also, both $g^{\phi}$ and $h^{\phi}$ don't depend on $v$, so the next Taylor term is null.

Using 
\begin{equation}
    sin^{2}\alpha\approx\frac{\bar{n}_2^{2}}{n_{1}^{2}}+2\frac{\bar{n}_2}{n_{1}}\sqrt{1-\frac{\bar{n}_2^{2}}{n_{1}^{2}}}(\bar{\alpha}_C-\alpha)
\end{equation}
and approximating at first order $v$ around $v=0$
\begin{equation}
    v\approx\sqrt{2\bar{n}_2\sqrt{n_{1}^{2}-\bar{n}_2^{2}}(\alpha-\bar{\alpha}_C)}
\end{equation}
The approximated sensitivities in phase are
\begin{equation}
    S_{p}^{\phi}\approx-\frac{\sqrt{2}n_{1}^{2}}{\bar{n}_2^{3/2}(n_{1}^{2}-\bar{n}_2^{2})^{3/4}}\frac{1}{\sqrt{\alpha-\alpha_{c}}}
\end{equation}
\begin{equation}
    S_{s}^{\phi}\approx-\frac{\sqrt{2}\sqrt{\bar{n}_2}}{(n_{1}^{2}-\bar{n}_2^{2})^{3/4}\sqrt{(\alpha-\bar{\alpha}_C)}}
\end{equation}




\bigskip


 \bibliographystyle{elsarticle-num} 
 \bibliography{sample}


\end{document}